\documentclass[review]{elsarticle}

\usepackage{amsmath,amssymb,bm}
\usepackage{physics}
\usepackage{color}
\usepackage{comment}

\usepackage{lineno,hyperref}
\usepackage{booktabs}
\usepackage{multirow}
\usepackage{xcolor}
\definecolor{darkgreen}{rgb}{0.0, 0.5, 0.0} 

\usepackage{tikz}

\usepackage{ulem}

\newcommand{\LA}{\mathop{\mathcal{L}}\nolimits}

\newcommand{\lM}{\lambda_{\rm M}}
\newcommand{\lm}{\lambda_{\rm m}}
\newcommand{\lSm}{\lambda_{\rm sm}}

\newcommand{\CA}{\mathop{\mathbb{C}}\nolimits}

\newcommand{\I}{\mathop{\mathbb{I}}\nolimits}

\newcommand{\bracket}[2]{\langle #1|#2 \rangle}

\newtheorem{Proposition}{Proposition}

\newtheorem{Lemma}{Lemma}

\newtheorem{Conjecture}{Conjecture}

\begin{document}

\begin{frontmatter}

\title{B\"ottcher-Wenzel inequality for weighted Frobenius norms and its application to quantum physics}

\author[1]{Aina Mayumi}
\ead{a.mayumi1441@gmail.com}
\author[1]{Gen Kimura*}
\ead{gen@shibaura-it.ac.jp}
\author[2]{Hiromichi Ohno}
\ead{h_ohno@shinshu-u.ac.jp}
\author[3]{Dariusz Chru\'sci\'nski}
\ead{darch@fizyka.umk.pl}
\address[1]{College of Systems Engineering and Science,
Shibaura Institute of Technology, Saitama 330-8570, Japan}
\address[2]{Department of Mathematics, Faculty of Engineering, Shinshu University,
4-17-1 Wakasato, Nagano 380-8553, Japan.}
\cortext[C]{Corresponding author}
\address[3]{Institute of Physics, Faculty of Physics, Astronomy and Informatics Nicolaus Copernicus University, Grudzi\c{a}dzka 5/7, 87--100 Toru\'n, Poland}

\begin{abstract}
By employing a weighted Frobenius norm with a positive matrix $\omega$, we introduce natural generalizations of the famous B\"ottcher-Wenzel (BW) inequality. 
Based on the combination of the weighted Frobenius norm $\|A\|_\omega := \sqrt{\tr(A^\ast A \omega)}$ and the standard Frobenius norm $\|A\| := \sqrt{\tr(A^\ast A)}$, there are exactly five possible generalizations, labeled (i) through (v), for the bounds on the norms of the commutator $[A,B]:= AB - BA$.
In this paper, we establish the tight bounds for cases (iii) and (v), and propose conjectures regarding the tight bounds for cases (i) and (ii). Additionally, the tight bound for case (iv) is derived as a corollary of case (i).
All these bounds (i)-(v) serve as generalizations of the BW inequality.
The conjectured bounds for cases (i) and (ii) (and thus also (iv)) are numerically supported for matrices up to size $n=15$. 
Proofs are provided for $n=2$ and certain special cases. 
Interestingly, we find applications of these bounds in quantum physics, particularly in the contexts of the uncertainty relation and open quantum dynamics.
\end{abstract}


\end{frontmatter}

\section{Introduction}

The seminal B\"ottcher-Wenzel (BW) inequality \cite{BOTTCHER2005216,vong2008proof,BOTTCHER20081864} provides the bound of the norm of the commutator $[A,B]:= AB - BA$ of $n \times n$ complex matrices $A,B \in M_n(\CA)$ in the form
\begin{align}\label{BW}
\|[A,B]\| \le \sqrt{2} \|A\| \|B\|,
\end{align}
where $\|A\| := \sqrt{\tr(A^\ast A)}$ is the Frobenius norm.
Here $\tr A$ and $A^\ast$ are the trace and the Hermitian conjugate of the matrix $A$, respectively. The bound \eqref{BW} is tight in the sense that there exist non-zero matrices $A,B$ that attain the equality. 
This inequality was then generalized in several directions, e.g., with Schatten $p$-norm, Ky Fan $(p,k)$ norm \cite{AUDENAERT20101126,WENZEL20101726,CHENG2015409}, or with the $q$-deformed commutator \cite{CHRUSCINSKI202295,CHRUSCINSKI2023158}. 

In this paper we consider generalizations of BW inequality by replacing the Frobenius norm defined in terms of the Hilbert Schmidt inner product $\bracket{A}{B} := \tr (A^\ast B)$, i.e. $\|A\|^2 = \bracket{A}{A}$, by $\|A\|_\omega^2 := \bracket{A}{A}_\omega$, where the new inner product is defined as follows $\bracket{A}{B}_\omega := \tr (A^*B\omega )$ and $\omega \in M_n(\CA)$ is a positive (definite) matrix. In what follows we call $\omega$-weighted Frobenius norm

\begin{equation}\label{eq:StateNorm}
\|A\|_\omega := \sqrt{\tr (A^\ast A \omega)}
\end{equation}
the {\it $\omega$-norm}. The $\omega$-norm satisfies the axioms of the norm and provides a generalization of the Frobenius norm when $\omega$ is the identity matrix $\I$. In the following discussion, $\omega$ is always assumed to be a positive matrix.

Depending on the combinations of the $\omega$-norm and the Frobenius norm, there are exactly six types of bounds to be considered, including the BW inequality itself\footnote{Note that the norm of the commutator $[A,B]$ exhibits symmetry between $A$ and $B$, hence interchanging the $\omega$-norm and the Frobenius norm for $A$ and $B$ in the right-hand side of \eqref{bounds} does not yield new types of bounds.}: For positive constants $c_i(\omega) \ (i=1,\ldots,6)$, dependent on $\omega$, it holds that, for any $A,B \in M_n(\CA)$,
\begin{subequations}\label{bounds}
\renewcommand{\theequation}{\theparentequation\roman{equation}}
	\begin{alignat}{3}
		&\rm{(i)}\ && \|[A,B]\|_\omega &&\le c_1(\omega) \|A\|_\omega \|B\|_\omega, \label{bounds1} \\
&\rm{(ii)}\ &&\|[A,B]\|_\omega &&\le c_2(\omega) \|A\|_\omega \|B\|, \label{bounds2} \\
	&\rm{(iii)}\  &&\|[A,B]\|_\omega &&\le c_3(\omega) \|A\| \|B\|, \label{bounds3} \\
&\rm{(iv)}\  &&\|[A,B]\| &&\le c_4(\omega) \|A\|_\omega \|B\|_\omega, \label{bounds4} \\
&\rm{(v)}\ &&\|[A,B]\| &&\le c_5(\omega) \|A\|_\omega \|B\|, \label{bounds5} \\
&\rm{(vi)}\  &&\|[A,B]\| &&\le c_6(\omega) \|A\| \|B\|\label{bounds6}.
	\end{alignat}
\end{subequations}
In particular, we are interested in the tightest bounds $\tilde{c_i}(\omega) \ (i=1,\ldots,6)$, which are the minimum values of $c_i(\omega)$ ($i=1,\ldots,6$) for which each inequality in Eqs. \eqref{bounds} is satisfied.
Alternatively, $\tilde{c_i}(\omega)$ can be characterized by the following optimization problems:
\begin{subequations}\label{tcopt}
	\renewcommand{\theequation}{\theparentequation\roman{equation}}
	\begin{align}
	\tilde{c_1}(\omega) &= \max_{A,B \neq 0 \in M_n(\CA)} \frac{\|[A,B]\|_\omega}{\|A\|_\omega \|B\|_\omega}, \label{tempi} \\
	\tilde{c_2}(\omega) &= \max_{A,B \neq 0 \in M_n(\CA)} \frac{\|[A,B]\|_\omega}{\|A\|_\omega \|B\|}, \label{tempii}\\
	\tilde{c_3}(\omega) &= \max_{A,B \neq 0 \in M_n(\CA)} \frac{\|[A,B]\|_\omega}{\|A\| \|B\|}, \\
	\tilde{c_4}(\omega) &= \max_{A,B \neq 0 \in M_n(\CA)} \frac{\|[A,B]\|}{\|A\|_\omega \|B\|_\omega}, \\
	\tilde{c_5}(\omega) &= \max_{A,B \neq 0 \in M_n(\CA)} \frac{\|[A,B]\|}{\|A\|_\omega \|B\|}, \\
	\tilde{c_6}(\omega) &= \max_{A,B \neq 0 \in M_n(\CA)} \frac{\|[A,B]\|}{\|A\| \|B\|}.
	\end{align}
\end{subequations}
Note that, given inequalities \eqref{bounds}, the tightness can be also shown by providing non-zero matrices $A,B$ such that the equalities are attained.

We remark that type (vi) corresponds to BW bound \eqref{BW}, so that we already know
\begin{align}\label{BWbdd}
	\tilde{c_6}(\omega) = \sqrt{2}.
\end{align}
In this paper, we first show that the tight bounds for cases (iii) and (v) are give by
\begin{align}\label{ctiiiAv}
\tilde{c_3}(\omega)= \sqrt{2\lM}, \ \tilde{c_5}(\omega)=\sqrt{\frac{2}{\lm}},
\end{align}
where $\lM$ and $\lm$ are the largest and the smallest eigenvalues of $\omega$.
Second, we give conjectures for cases (i) and (ii) that the tight bounds are given respectively by
\begin{align}\label{tc1}
	\tilde{c_1}(\omega) = \sqrt{\frac{\lm + \lSm}{\lm \lSm}}
\end{align}
and
\begin{align}\label{tc2}
	\tilde{c_2}(\omega) = \sqrt{\frac{\lm + \lM}{\lm}},
\end{align}
where $\lSm$ is the second smallest eigenvalue of $\omega$.
Third, the tight bound for case (iv) is then given, as a corollary of conjecture \eqref{tc1}, by
\begin{align}\label{tc4}
	\tilde{c_4}(\omega) = \sqrt{\frac{\lm + \lSm}{\lm^2 \lSm}}.
\end{align}
Note that all these tight bounds for cases (i)-(v) generalize the BW inequality by setting $\omega = \I$, since the eigenvalues of $\I$ are all 1, allowing \eqref{ctiiiAv}, \eqref{tc1}, \eqref{tc2}, and \eqref{tc4} to recover the BW bound \eqref{BWbdd}.

To summarize, the logical relations between the tight bounds for cases (i)-(vi) are as follows:
\begin{figure}[h]
	\centering
	\includegraphics[width = 9cm]{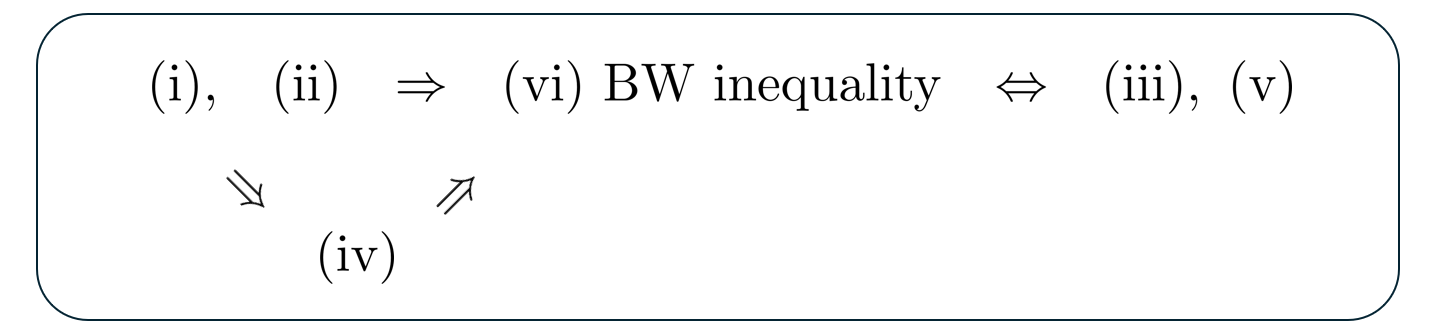}
\end{figure}

\noindent In this regard, it becomes evidence that the verification of conjectures (i) and (ii) holds considerable importance.
Furthermore, besides the intrinsic mathematical interest in the norm bounds of the commutator, we find that there are direct applications of both (i) and (ii) in the filed of quantum physics.
By applying a quantum state, i.e., a density operator as $\omega$, we demonstrate that the conjectured bounds for type (i) and type (ii) respectively introduce a novel uncertainty relation between observables $A$ and $B$, and impose a non-trivial constraint on relaxation rates in the general quantum Markovian dynamics.

In the following discussion, we will frequently use the Dirac notation, as commonly used in the field of quantum physics\footnote{For a detailed explanation of the Dirac notation, see, e.g., \cite{Hayashi2015}.}: a vector in $\CA^n$ is denoted by a ket vector, e.g., $\ket{\psi}$;
For vectors $\ket{\psi}=(x_i)_{i=1}^n,\ket{\phi}=(y_i)_{i=1}^n \in \CA^n$, the symbol $\bracket{\psi}{\phi} := \sum_i \overline{x_i} y_i$ denote the (complex Euclidean) inner product, while the symbol $\ketbra{\psi}{\phi} :=[x_i \overline{y}_j]_{i,j}$ denotes the matrix in $M_n(\CA)$ such that its action to a vector is given by: $\ketbra{\psi}{\phi} \ket{\xi} = \bracket{\phi}{\xi} \ket{\psi}$.

The structure of this paper is as follows: In Sec.~\ref{sec:Bdd} (and also in Appendix), we explore several generalizations of the BW inequality, each characterized by a combination of the $\omega$-norm and the Frobenius norm. Sec.~\ref{sec:App} demonstrates the practical applications of our conjectures within the realm of quantum physics. The paper concludes with Sec.~\ref{sec:conc}.

\section{Six types of bounds of the commutators with respect to $\omega$-norm}\label{sec:Bdd}

To derive bounds for types (i) through (v) in \eqref{bounds}, we start with a simple observation about the relationship between the $\omega$-norm and the Frobenius norm. In what follows, we denote by $\lambda_i > 0$ $(i=1,\ldots,n)$ the eigenvalues of a positive matrix $\omega \in M_n(\mathbb{C})$, arranged in ascending order: $\lambda_1 \le \lambda_2 \le \cdots \le \lambda_n$, so that 
$$
\lm = \lambda_1, \lSm = \lambda_2 \ {\rm and} \ \lM = \lambda_n. 
$$
The corresponding unit eigenvectors are denoted by $\ket{\lambda_i}$ $(i=1,\ldots,n)$, i.e., $\omega \ket{\lambda_i} = \lambda_i \ket{\lambda_i}$ and $\bracket{\lambda_i}{\lambda_i}=1$.
Considering the expressions $\|A\|^2_\omega = \tr (A^\ast A \omega) = \sum_{i=1}^n \lambda_i \bracket{\lambda_i}{A^\ast A \lambda_i}$ and $ \|A\|^2 = \tr (A^\ast A) = \sum_{i=1}^n \bracket{\lambda_i}{A^\ast A \lambda_i}$, it follows that
\begin{align}\label{minOmegamax}
	\lm \|A\|^2 \le \|A\|^2_\omega  \le \lM\|A\|^2.
\end{align}
By utilizing these inequalities, we derive that
$$
\|[A,B]\|_\omega^2 \le \lM \|[A,B]\|^2 \le 2\lM \| A \|^2 \| B \|^2 \le \frac{2\lM}{\lm} \| A \|_\omega^2 \| B \|^2 \le \frac{2\lM}{\lm^2} \| A \|_\omega^2 \| B \|_\omega^2,
$$
where we have used the BW inequality \eqref{BW} to estimate the second inequality.
Now the inequalities clearly imply bounds (i)-(v) in \eqref{bounds} for
\begin{align}\label{ci}
	c_1(\omega)=\frac{\sqrt{2\lM}}{\lm}, c_2(\omega)=\sqrt{\frac{2\lM}{\lm}}, c_3(\omega)= \sqrt{2\lM}, c_4(\omega)=\frac{\sqrt{2}}{\lm}, c_5(\omega) = \sqrt{\frac{2}{\lm}}.
\end{align}
It turns out that this simple observation yields the tight bounds for cases (iii) and (v), since there are non-zero matrices $A,B$ that attain the equalities of \eqref{bounds3} and \eqref{bounds5} with $c_3(\omega)$ and $c_5(\omega)$ in \eqref{ci}.
Examples of such matrices include, for instance, $A=\ketbra{\lambda_1}{\lambda_n}$ and $ B=\frac{1}{\sqrt{2}}(\ketbra{\lambda_n}{\lambda_n}-\ketbra{\lambda_1}{\lambda_1})$ for \eqref{bounds3} and $A=\ketbra{\lambda_n}{\lambda_1}$ and $B=\ketbra{\lambda_1}{\lambda_n}$ for \eqref{bounds5}.

\begin{figure}[tbh]
	\includegraphics[width=\textwidth]{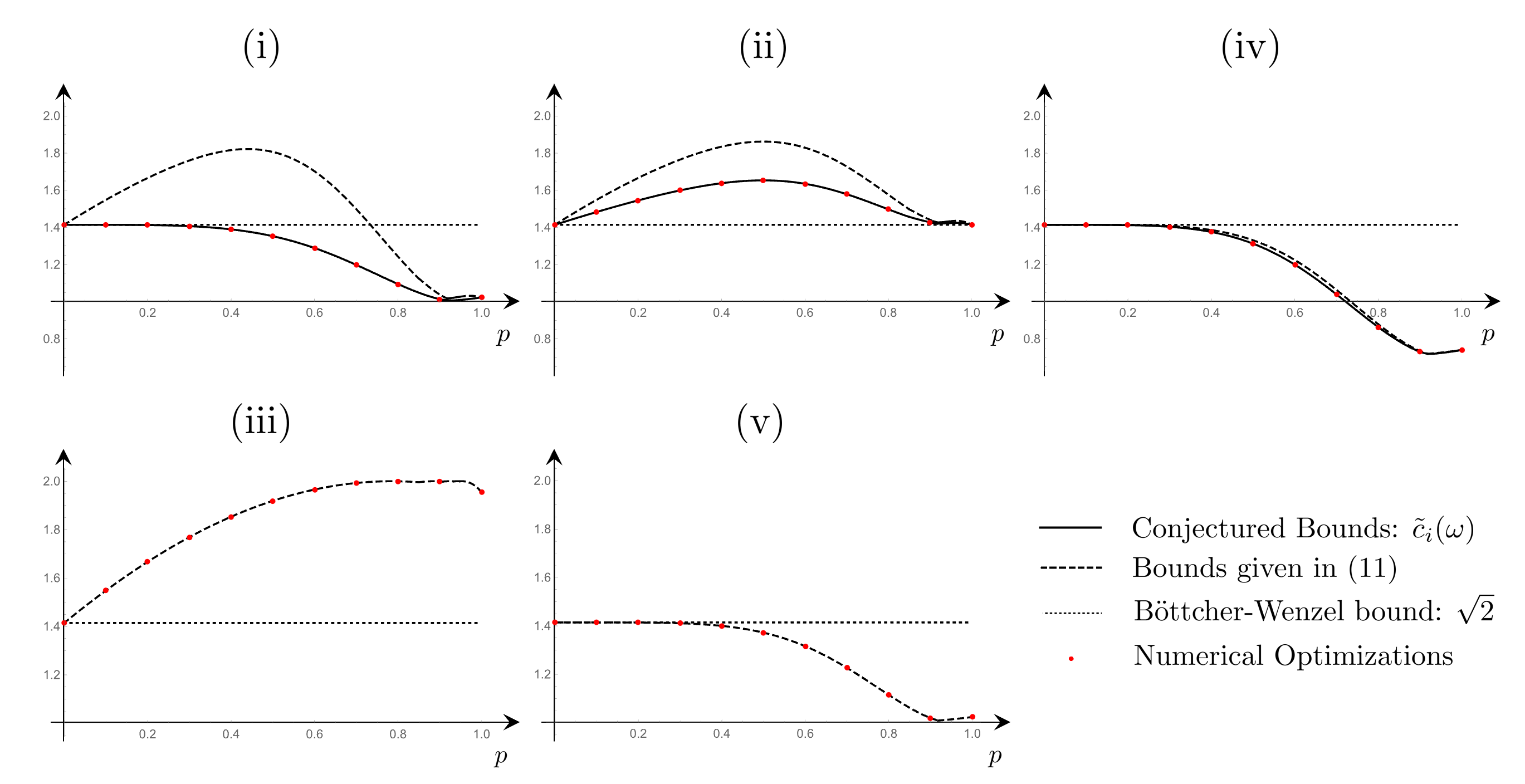}
	\caption{As an illustration, using for a $5\times 5$ positive matrix $\omega(p) = {\rm diag}[\sin(2p),\sin(2p^2),\sin(2p^3),\sin(2p^4),\sin(2p^5)]$ for $p \in (0,1]$, we compare numerical optimizations (dotted points/red online) with our conjectures.
		In graphs (i), (ii), (iv), conjectured bounds \eqref{tc1}, \eqref{tc2}, and \eqref{tc4} are plotted with solid lines, while bounds given in \eqref{ci} are plotted with dashed lines. For references, in graphs (iii) and (v), the tight bounds given in \eqref{ci} are plotted with dashed lines.
	}\label{num}
\end{figure}

Put differently, we have shown the following:
\begin{Proposition} Let $\omega \in M_n(\CA)$ be a positive matrix. For any matrices $A,B \in M_n(\CA)$,
	\begin{align}
		\|[A,B]\|_\omega \le \sqrt{2\lM} \|A\| \|B\|
	\end{align}
	and
	\begin{align}
		\|[A,B]\| \le \sqrt{\frac{2}{\lm}} \|A\|_\omega \|B\|,
	\end{align}
where $\lM,\lm $ are the largest and the smallest eigenvalues of $\omega$. Both bounds are tight, i.e., there are non-zero matrices $A$ and $B$ that attain the equalities.
\end{Proposition}
Meanwhile, we have conducted numerical optimizations of \eqref{tcopt} for cases (i), (ii), and (iv), and the results suggest that the bounds $c_1(\omega),c_2(\omega)$ and $c_4(\omega)$ in \eqref{ci} are not tight (See dashed lines in Fig.~\ref{num}).
Instead, we conjecture for the tight bounds as specified in \eqref{tc1}, \eqref{tc2}, and \eqref{tc4}.
In other words, our conjectures for cases (i), (ii) and (iv) read:
\begin{Conjecture}\label{Conj} For any matrices $A,B \in M_n(\CA)$,
	\begin{align}\label{Conji}
		\|[A,B]\|_\omega \le \sqrt{\frac{\lm + \lSm}{\lm \lSm}} \|A\|_\omega \|B\|_\omega,
	\end{align}
	\begin{align}\label{Conjii}
		\|[A,B]\|_\omega \le \sqrt{\frac{\lm + \lM}{\lm}} \|A\|_\omega \|B\|,
	\end{align}
and (as a corollary of \eqref{Conji})
	\begin{align}\label{Conjiv}
	\|[A,B]\| \le \sqrt{\frac{\lm + \lSm}{\lm^2 \lSm}} \|A\|_\omega \|B\|_\omega.
    \end{align}
All bounds are tight, i.e., there are non-zero matrices $A$ and $B$ that attain the equalities.
\end{Conjecture}

Some remarks are in order: First, these conjectures are strongly supported by numerical evidence: For randomly generated positive matrices $\omega$, we have confirmed that numerical optimizations of \eqref{tcopt} for cases (i), (ii), and (iv) perfectly match the conjectured bounds  \eqref{tc1}, \eqref{tc2}, and \eqref{tc4} up to size $n = 15$.  For illustration, consider a one parameterized matrix
$$
\omega(p) = \text{diag}[\sin(2p), \sin(2p^2), \sin(2p^3), \sin(2p^4), \sin(2p^5)] \in M_5(\mathbb{C})
$$
with parameter $p \in (0,1]$ as a positive matrix $\omega$. (This matrix is just one example of $\omega$, but note that due to the positivity of $\omega$, it can always be diagonalized, so considering $\omega$ as a diagonal matrix does not result in any loss of generality.) Figure~\ref{num} illustrates the comparisons between numerical optimizations (dotted points, red in the online version) and the conjectured bounds (solid lines), as well as the loose bounds (dashed lines) given in \eqref{ci}.
\noindent Second, one can find non-zero matrices $A,B$ that attain equalities in \eqref{Conji}, \eqref{Conjii}, and \eqref{Conjiv}:
For type (i), let, for instance, $A = \ketbra{\lm}{\lSm}$ and $B = A^\ast$.
Then, it holds that $\|[A,B]\|^2_\omega = \lm + \lSm, \|A\|^2_\omega = \lSm, \|B\|^2_\omega = \lm$ and the equality in \eqref{Conji} is attained. Similarly, attainability can be shown by setting, for type (ii), $A = \ketbra{\lM}{\lm}$, $B = A^\ast$, and for type (iv), $A=\lm \ketbra{\lSm}{\lSm} -\lSm \ketbra{\lm}{\lm},
B=\ketbra{\lSm}{\lm}$, respectively.
These facts imply that the proofs of the part concerning the tightness of the bounds are complete.
Third, \eqref{Conjiv} can be obtained as a corollary of \eqref{Conji} by using the left inequality in \eqref{minOmegamax} for the commutator part.
Forth, we provide proofs of the conjectures for some special cases (see \ref{app:Acom}); At the end of this section, we give proofs for case $n=2$.
Last but not least, we emphasize that the forms of our conjectures are far from trivial. For instance, the fact that the tight bounds depend solely on the eigenvalues of $\omega$, and more specifically, on its largest, smallest, or second smallest eigenvalues, as presented in forms \eqref{tc1}, \eqref{tc2}, \eqref{tc4}, was completely beyond our initial expectations. We arrived at these conjectured forms only after conducting numerous numerical experiments and trials.

\subsection{Proofs of Conjecture \ref{Conj} for $n=2$}

Here we prove inequalities \eqref{Conji} and \eqref{Conjii}, hence Conjecture \ref{Conj}, for the case $n=2$.
Before addressing the specific case of $n=2$, we begin with several general observations applicable to matrices of any size $n$.
First, in order to prove the conjecture for types (i) and (ii) (indeed also for all other cases), we note that, without loss of generality, $\omega$ can be assumed to be a diagonal matrix with unit trace.
This follows from the facts that, for any unitary matrix $U$, $\|U A U^\ast \|_{U \omega U^\ast} = \|A \|_{\omega}$ (thus $\|U A U^\ast \| = \|A \|$) and $[UAU^\ast,UBU^\ast] = U[A,B]U^\ast$.
Moreover, for any positive constant $p$, $\|A \|_{p\omega} = \sqrt{p} \|A \|_{\omega}$, and therefore $\sqrt{p}\tilde{c_1}(p\omega) =
\tilde{c_1}(\omega)$ and $\tilde{c_2}(p\omega) = \tilde{c_2}(\omega)$ for \eqref{tc1} and \eqref{tc2}. 
Note that these are also true for \eqref{tempi} and \eqref{tempii}.

Second, note that $M_n(\mathbb{C})$ constitutes an inner product space with respect to the $\omega$-inner product, defined as $\bracket{A}{B}_\omega := \tr (A^* B \omega)$, where the $\omega$-norm is induced from it. As the identity matrix is a unit vector if $\omega$ has a unit trace, we have
\begin{align}\label{nAI}
	\|A\|^2_\omega \ge \|A - \bracket{I}{A}_\omega I \|^2_\omega = \|A - \tr(A\omega) I \|^2_\omega.
\end{align}
Applying $\omega = \I/n$, we also have
\begin{align}\label{nAIF}
	\|A\|^2 \ge  \|A - \frac{\tr(A)}{n} I \|^2.
\end{align}

Considering these observations, we now proceed to prove \eqref{Conji} and \eqref{Conjii} in the case where $n=2$.
In what follows, let $\omega ={\rm diag}[\lambda_1, \lambda_2]$ with $\lambda_1 + \lambda_2 = 1$ and $\lambda_1 \le \lambda_2$, and $A=[a_{ij}]_{i,j}, B=[b_{ij}]_{i,j} \in M_2({\mathbb C})$.

[Proof of \eqref{Conji}] Note that inequality \eqref{nAI} along with the fact $[A + \alpha \I,B+ \beta \I ] = [A, B]$ for any $\alpha,\beta \in {\mathbb C}$ allows us to assume that $A$ and $B$ take the forms $A - \tr(A\omega) I \ {\rm and} \ B - \tr(B\omega) I$, respectively. This is equivalent to assuming $a_{22} = -\frac{\lambda_1}{\lambda_2}a_{11}$ and $b_{22} = -\frac{\lambda_1}{\lambda_2}b_{11}$. 
The direct computation then yields the identity
\begin{align}\label{id14}
	\frac{\lm + \lSm}{\lm\lSm}
	\|A\|^2_\omega  \|B\|^2_\omega - \|[A,B]\|^2_\omega = \left| \frac{1}{\lambda_2}\sqrt{\frac{\lambda_1}{\lambda_2}} a_{11} \bar{b}_{11} +
	\sqrt{\frac{\lambda_1}{\lambda_2}} a_{21}\bar{b}_{21}
	+\sqrt{\frac{\lambda_2}{\lambda_1}}a_{12}\bar{b}_{12} \right|^2,
\end{align}
where we have used $\lm = \lambda_1,\lSm = \lambda_2$ and $\lambda_1 + \lambda_2 = 1$. 
This clearly implies inequality \eqref{Conji}. 

[Proof of \eqref{Conjii}] We may assume $B$ to take the form $B - \frac{\tr(B)}{n} I$ by using \eqref{nAIF}, while $A$ retains the form $A - \tr(A\omega) I$ as above. Thus, $a_{22} = -\frac{\lambda_1}{\lambda_2}a_{11}$ and $b_{22} = -b_{11}$.
Noting that $\lm = \lambda_1,\lM = \lambda_2$, $\lambda_1 + \lambda_2 = 1$ and $\lambda_2 \ge \frac{1}{2} \ge \lambda_1$, we have
\begin{align*}
	&\frac{\lM + \lm}{\lm}
	\|A\|^2_\omega  \|B\|^2 - \|[A,B]\|^2_\omega = \frac{2}{\lambda_2}  |a_{11}|^2 |b_{11}|^2 + |a_{21}|^2 |b_{21}|^2 + \frac{\lambda_2}{\lambda_1}  |a_{12}|^2 |b_{12}|^2 \\
	&\quad +  \frac{\lambda_2-\lambda_1}{\lambda_1}  |a_{12}|^2 |b_{21}|^2 + 2(1-2\lambda_1)|a_{21}|^2 |b_{11}|^2+ \frac{\lambda_2-\lambda_1}{\lambda^2_2}  |a_{11}|^2 |b_{21}|^2 + \frac{2\lambda_2(1-2\lambda_1)}{\lambda_1} |a_{12}|^2 |b_{11}|^2 \\
	& \quad +
	2{\rm Re} a_{12}\bar{a}_{21} \bar{b}_{12}b_{21}
	+\frac{4\lambda_1}{\lambda_2}{\rm Re} a_{11}\bar{a}_{21} \bar{b}_{11} b_{21}
	+4{\rm Re} a_{11}\bar{a}_{12} \bar{b}_{11} b_{12} \\
	&\ge \frac{2}{\lambda_2}  |a_{11}|^2 |b_{11}|^2 + |a_{21}|^2 |b_{21}|^2 + \frac{\lambda_2}{\lambda_1}  |a_{12}|^2 |b_{12}|^2 \\
	& \quad +2{\rm Re} a_{12}\bar{a}_{21} \bar{b}_{12}b_{21} +\frac{4\lambda_1}{\lambda_2}{\rm Re} a_{11}\bar{a}_{21} \bar{b}_{11} b_{21} +4{\rm Re} a_{11}\bar{a}_{12} \bar{b}_{11} b_{12} \\
	&\ge \frac{4\lambda_1}{\lambda_2}|a_{11}|^2 |b_{11}|^2 + \frac{\lambda_1}{\lambda_2}|a_{21}|^2 |b_{21}|^2 + \frac{\lambda_2}{\lambda_1}  |a_{12}|^2 |b_{12}|^2 \\
	& \quad +2{\rm Re} a_{12}\bar{a}_{21} \bar{b}_{12}b_{21} +\frac{4\lambda_1}{\lambda_2}{\rm Re} a_{11}\bar{a}_{21} \bar{b}_{11} b_{21} +4{\rm Re} a_{11}\bar{a}_{12} \bar{b}_{11} b_{12} \\
	&= \left| 2\sqrt{\frac{\lambda_1}{\lambda_2}} a_{11} \bar{b}_{11} +
	\sqrt{\frac{\lambda_1}{\lambda_2}} a_{21}\bar{b}_{21}
	+\sqrt{\frac{\lambda_2}{\lambda_1}}a_{12}\bar{b}_{12} \right|^2
\end{align*}
which implies inequality \eqref{Conjii}.  

\section{Applications to quantum physics}\label{sec:App}

This section is dedicated to exploring how both conjectures \eqref{Conji} and \eqref{Conjii} can be effectively applied in the realm of quantum physics, demonstrating their utility and relevance.
In Sec.~\ref{appUR}, we will see that conjecture \eqref{Conji} unveils a new type of uncertainty relation for quantum observables, and in Sec.~\ref{appRC}, we will discuss how conjecture \eqref{Conjii} provides a constraint on relaxation rates in quantum Markovian dynamics.

\subsection{Application of Conjecture \eqref{Conji}}\label{appUR}

Our primary motivation for introducing the $\omega$-norm was to apply BW inequality to the uncertainty relations in quantum physics.
In the context of a $n$-level quantum system $\CA^n$, physical quantities (observables) are represented by Hermitian matrices $A, B$, where a quantum state is represented by a density matrix $\rho$, a positive semidefinite matrix with a unit trace.
Under a state $\rho$, the expectation value of $A$ is given by $E(A)_\rho = \tr \rho A$, and the variance of $A$ is given by
\begin{align}\label{VAq}
	 V(A)_\rho = \tr \rho A^2 - (\tr \rho A)^2 = \tr \rho(A- (\tr A\rho)\I )^2.
\end{align}
The famous Robertson uncertainty relation is
\begin{equation}\label{RU}
	V(A)_\rho V(B)_\rho \ge \frac{1}{4}|\tr \rho [A,B]|^2.
\end{equation}
Therefore, in the field of quantum physics, it is understood that the non-commutativity of physical quantities results in a trade-off between their uncertainties.

On the other hand, it is straightforward to observe that the BW inequality offers a comparable uncertainty relation for observables when a quantum system is in the maximally mixed state, denoted as $\rho_{\max} = \frac{\I}{n}$.
Specifically, for this state, the variances can be represented through the Frobenius norm by
$$
V(A)_{\rho_{\max}} = \frac{1}{n} \|A - \frac{\tr A}{n} \I \|^2, V(B)_{\rho_{\max}} = \frac{1}{n} \|B - \frac{\tr B}{n} \I \|^2.
$$
Thus, noting that $[A - \frac{\tr A}{n} \I,B - \frac{\tr B}{n} \I] = [A,B]$, the BW inequality provides the following uncertainty relation:
\begin{align}\label{unBW}
	V(A)_{\rho_{\max}}V(B)_{\rho_{\max}} \ge \frac{1}{2n^2} \|[A,B]\|^2. 
\end{align}
Applying the maximally mixed state to the Robertson relation \eqref{RU}, the bound on the right-hand side is given by $\frac{1}{4n^2}|\tr [A,B]|^2$. Consequently, Eq. \eqref{unBW} yields a similar, yet distinct, uncertainty relation. As an example, consider $A = \sigma_x$ and $B = \sigma_y$, where $\sigma_x = \begin{pmatrix} 0 & 1 \\ 1 & 0 \end{pmatrix}$, $\sigma_y = \begin{pmatrix} 0 & -i \\ i & 0 \end{pmatrix}$, and $\sigma_z = \begin{pmatrix} 1 & 0 \\ 0 & -1 \end{pmatrix}$ are the Pauli matrices for a two-level quantum system. Since $[\sigma_x,\sigma_y] = 2i\sigma_z$ and $\tr \sigma_z = 0$ the Robertson bound yields a trivial bound $0$, while the bound in \eqref{unBW} is $1$.

\begin{figure}
	\includegraphics[width=\textwidth]{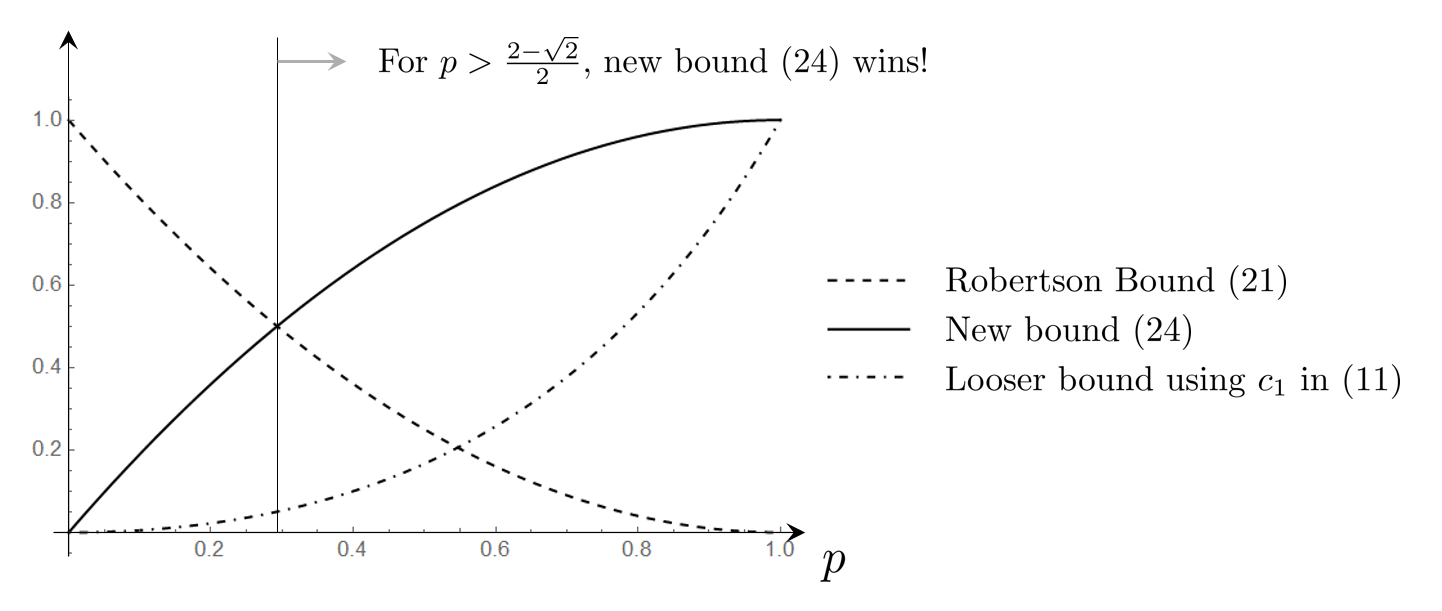}
	\caption{The Robertson bound \eqref{RU} and our bound \eqref{newU} between observables $\sigma_x$ and $\sigma_y$ under a qubit state $\rho(p) = p \frac{\I}{2} + (1-p) \ketbra{0}{0} \ (p \in (0,1])$ are plotted with dashed and solid lines, respectively. For comparison, the looser bound, using $c_1$ in \eqref{ci} is also plotted with dot-dashed line.
	}\label{fig2}
\end{figure}

Unfortunately, the relation \eqref{unBW} is only valid for the maximally mixed state. To extend this to an arbitrary state, the $\omega$-norm becomes essential: By defining $\tilde{A} := A - (\tr \rho A) \I$ and $\tilde{B} := B - (\tr \rho B) \I$, the variance of $A$ and $B$ under a state $\rho$ can be represented by
\begin{align}\label{VAB}
	V(A)_\rho = \|\tilde{A}\|^2_\rho, V(B)_\rho = \|\tilde{B}\|^2_\rho.
\end{align}	
Thus, type (i) bound (Eq. \eqref{bounds1}) --- as well as type (iv) --- introduces another uncertainty principle that is valid for any quantum state $\rho$.
Notably, conjecture \eqref{Conji} leads to a novel form of the uncertainty relation: For any faithful quantum state $\rho$ (i.e., a positive definite matrix with unit trace) and for any observables $A$ and $B$,
\begin{equation}\label{newU}
	V(A)_\rho V(B)_\rho \ge \frac{\lm \lSm}{\lm + \lSm} \|[A,B]\|^2_\rho,
\end{equation}
where $\lm$ and $\lSm$ denote the smallest and the second smallest eigenvalues of $\rho$, respectively.
For comparison, if we use a bound $c_1$ (which is already proven) given in \eqref{ci}, we have the looser uncertainty relation: 
\begin{align}\label{newU2}
V(A)_\rho V(B)_\rho \ge \frac{\lm^2}{2\lM} \| [A,B]\|^2_\rho.	
\end{align} 
As illustrations, let's consider $A = \sigma_x$ and $B = \sigma_y$ again and a quantum state which is a probabilistic mixture of the maximally mixed state and the eigenstate $\ket{0}$ of $\sigma_z$ corresponding to the eigenvalue $1$, i.e., $\rho(p) = p \frac{\I}{2} + (1-p) \ketbra{0}{0} \ (p \in (0,1])$. 
Simple computations shows that the bounds in \eqref{RU}, \eqref{newU} and \eqref{newU2} are given respectively by $(1-p)^2, p(2-p)$, and $\frac{p^2}{2-p}$. 
Therefore, the bound \eqref{newU} outperforms the Robertson bound \eqref{RU} if $p > (2-\sqrt{2})/2 \simeq 0.293$ (See Fig.~\ref{fig2}). 
Interestingly, even the looser bound \eqref{newU2} surpasses the Robertson bound if $p \gtrsim 0.547$. 
In \cite{Mayumi_2023}, we conduct a comprehensive comparison of these new uncertainty relations with the standard ones, including not only the Robertson relation but also the Schrödinger relation.
It is demonstrated that the bounds in \eqref{newU} and \eqref{newU2} unveil entirely new types of trade-offs in quantum uncertainty that were previously undetected by the standard relations, particularly in cases where the state exhibits a higher degree of mixedness.

\subsection{Application of Conjecture \eqref{Conjii}}\label{appRC}

Although conjecture \eqref{Conjii} employs an asymmetric combination of the $\omega$-norm and the Frobenius norm, making it appear artificial and pedantic, it finds direct application in the field of open quantum dynamics.

In the field of open quantum system, it is widely recognized that general quantum Markovian dynamics is described by a completely positive (CP) dynamical semigroup:
$$
\frac{d \rho_t}{dt} = \LA \rho_t,
$$
where $\rho_t$ is a quantum state at time $t$ and $\LA$ is the generator of time evolution.
One of the seminal works in this area is the representation theorem of the generator: $\LA$ qualifies as a generator for a CP dynamical semigroup if and only if it has the form
\begin{equation}\label{eq:GKLS}
	\LA(\rho) = -i[H,\rho] + \frac{1}{2}\sum_k \gamma_k(2 L_k \rho L_k^\dagger - L_k^\dagger L_k \rho -\rho L_k^\dagger L_k ),
\end{equation}
where $\gamma_k \ge 0$ and jump operators $L_k$ are normalized as $\|L_k\| = 1$.
Since this was independently discovered by Gorini, Kossakowski, Sudarshan \cite{Gorini1976CompletelyPositive} and by Lindblad \cite{Lindblad1976OnTheGenerators}, the generator is now referred to as the GKLS (or GKSL) generator (See also \cite{Chruscinski2017BriefHistoryGKLS}).

Letting $\LA^\ddagger$ be the dual of $\LA$ via $\tr( X\LA(Y)) = \tr (\LA^\ddagger (X) Y)$ and $\ell_\alpha$ and $Y_\alpha$ be an eigenvalue and an eigenvector of $\LA^\ddagger$: $\LA^\ddagger (Y_\alpha) = \ell_\alpha Y_\alpha$. Note that a relaxation rate $\Gamma_\alpha$ is given by the real part of the eigenvalue $\ell_\alpha$. Let $\omega$ be a faithful stationary state, i.e., $\LA(\omega) = 0$.
In \cite{PhysRevLett.127.050401}, a relaxation rate $\Gamma_\alpha$ for any GKLS generator has been characterized by
\begin{equation}\label{key}
	\Gamma_\alpha = \frac{1}{2 \|Y_\alpha\|^2_\omega} \sum_k \gamma_k \|[L_k,Y_\alpha]\|^2_\omega.
\end{equation}
Now, by applying conjecture \eqref{Conjii} and recalling that $\|L_k\| = 1$, we have
$$
\Gamma_\alpha \le \frac{1}{2}\Bigl(1 + \frac{\lM}{\lm}\Bigr) \sum_k \gamma_k,
$$
where $\lM$ and $\lm$ are the maximal and the minimal eigenvalues of the stationary state $\omega$.
On the other hand, as we have $\sum_k \gamma_k = \frac{1}{n} \sum_{\beta=1}^{n^2-1} \Gamma_\beta$ \cite{WolfCirac2008,KimuraAjisakaWatanabe2017}, we obtain the following constraints on relaxation rates for GKLS master equation:
\begin{equation}\label{eq:constr}
		\Gamma_\alpha \le \frac{1}{2n}\Bigl(1 + \frac{\lM}{\lm}\Bigr) \sum_{\beta=1}^{n^2-1} \Gamma_\beta.
\end{equation}
Note that relaxation rates are observable in experiments, the above constraint can be directly tested in experimental setups.

\section{Concluding remarks}\label{sec:conc}

In this paper, we investigated generalizations of the BW inequality, utilizing combinations of the $\omega$-norm and the Frobenius norm. There are six types of generalization, types (i)-(vi), with type (vi) being the BW inequality itself.
We have established the tight bounds for types (iii) and (v), and offered conjectures for the tight bounds of types (i) and (ii), with type (iv) emerging as a corollary of type (i).
Both conjectures for types (i) and (ii) are backed by numerical evidence and have been validated for matrices of size $n=2$ and certain special cases for general $n$.
Additionally, we showed that both types (i) and (ii) have direct applications to quantum physics.
The type (i) bound introduces a novel uncertainty relation that reveals a previously undetected trade-off in the uncertainties between non-commuting observables.
The type (ii) bound imposes a significant constraint on relaxation rates within general quantum Markovian dynamics, offering a directly testable prediction in experimental settings.

We hope that our results open a new direction of generalizations of BW inequality, enriching both the mathematical and physical perspectives on the subject.

\section*{Acknowledgments}
AM and GK would like to thank Jaeha Lee for the insightful discussions on the uncertainty relations.
HO was supported by JSPS KAKENHI Grant Number 23K03147 and DC was supported by the Polish National Science Center project No. 2018/30/A/ST2/00837.

\appendix

\section{Proofs of conjectures for special cases}\label{app:Acom}

In this appendix, we give proofs of \eqref{Conji} and \eqref{Conjii} in some restricted cases.

\bigskip

\noindent [The case where $A$ is normal and commutes with $\omega$]
\bigskip

Here we give proofs of \eqref{Conji} and \eqref{Conjii} in the case where $A$ is normal and commutes with $\omega$.
In this case, we can assume that both $A$ and $\omega$ are diagonal: Let
$A={\rm diag}[a_1, \ldots, a_n] \ (a_i \in {\mathbb C})$ and $\omega ={\rm diag} [\lambda_1, \ldots , \lambda_n]$
$(\lambda_1 \le \lambda_2 \le \cdots \le \lambda_n)$ and $B = [b_{ij}] \in M_n({\mathbb C})$.

Now, a direct computation yields:
\begin{align}
	&\frac{\lm + \lSm}{\lm\lSm}\|A\|^2_\omega \|B\|^2_\omega -
	\|[A,B]\|^2_\omega \nonumber \\
	&\quad =
	\sum_{i,j} |b_{ij}|^2 \lambda_j \left(
	\frac{\lambda_1 + \lambda_2}{\lambda_1\lambda_2}
	\sum_k |a_k|^2 \lambda_k - |a_i -a_j|^2 \right) \nonumber \\
	&\quad \ge
	\sum_{i,j} |b_{ij}|^2 \lambda_j  \left(
	\left(\frac{\lambda_i}{\lambda_1} + \frac{\lambda_i}{\lambda_2}-1 \right) |a_i|^2
	+ 2 {\rm Re}(a_i \bar{a}_j )
	+ \left(\frac{\lambda_j}{\lambda_1} + \frac{\lambda_j}{\lambda_2}-1 \right) |a_j|^2\right).
	\label{eq:i-A}
\end{align}
Note that $T(i) := \frac{\lambda_i}{\lambda_1} + \frac{\lambda_i}{\lambda_2}-1$ is larger than or equal to $1$ when $i\neq 1$, and satisfies $
T(i)T(1) \ge 1$ for any $2 \le i \le n$. This clearly shows that \eqref{eq:i-A} is non-negative,  confirming \eqref{Conji}.

Similarly, \eqref{Conjii} can also be demonstrated as
\begin{align}
	&\frac{\lM + \lm}{\lm}\|A\|^2_\omega \|B\|^2 -
	\|[A,B]\|^2_\omega \nonumber\\
	&\quad =
	\sum_{i,j} |b_{ij}|^2 \left(
	\frac{\lambda_{n} + \lambda_{\rm 1}}{\lambda_{\rm 1}}
	\sum_k |a_k|^2 \lambda_k - |a_i -a_j|^2 \lambda_j \right)\nonumber\\
	&\quad \ge
	\sum_{i,j} |b_{ij}|^2 \left(
	\left(\lambda_i + \frac{\lambda_{n}}{\lambda_{\rm 1}}\lambda_i - \lambda_j\right) |a_i|^2
	+ 2 {\rm Re}(a_i \bar{a}_j \lambda_j)
	+ \frac{\lambda_{n}}{\lambda_{\rm 1}} \lambda_j|a_j|^2\right)\nonumber\\
	&\quad \ge
	\sum_{i,j} |b_{ij}|^2 \left( \lambda_i |a_i|^2
	+ 2 {\rm Re}(a_i \bar{a}_j \lambda_j)
	+ \frac{\lambda_j}{\lambda_{i}} \lambda_j|a_j|^2\right)\nonumber\\
	& \quad =
	\sum_{i,j} |b_{ij}|^2 \left| \sqrt{\lambda_i} a_i + \frac{\lambda_j}{\sqrt{\lambda_i}} a_j \right|^2 \ge 0.
	\label{eq:ii-A}
\end{align}

\bigskip

\noindent [The case where $B$ commutes with $\omega$]

\bigskip

When matrix $B$ commutes with $\omega$, conjecture \eqref{Conjii} can be easy to show. Moreover, the tight bound in this case coincides with BW bound, i.e., $\|[A,B]\|_\omega\le \sqrt{2}\|A\|_\omega\|B\|$.
To show this, we use the relation:
\begin{equation}\label{eq:RNorm}
	\|A\|_\omega = \|A \sqrt{\omega}\|,
\end{equation}
which follows from the cyclic property of trace as $\|A\|^2_\omega = \tr A^\ast A \omega = \tr \sqrt{\omega} A^\ast A \sqrt{\omega} = \tr (A \sqrt{\omega})^\ast A \sqrt{\omega} = \|A \sqrt{\omega}\|^2$.
Given that $B$ commutes with $\omega$, and therefore with $\sqrt{\omega}$, the relation $[B,A]\sqrt{\omega} = [B,A\sqrt{\omega}]$ is satisfied.
By using \eqref{eq:RNorm} and BW inequality, we have
$$
\| [A,B] \|^2_\omega = \| [B,A] \|^2_\omega = \|[B,A]\sqrt{\omega}\|^2 = \|[B,A\sqrt{\omega}]\|^2 \le \sqrt{2} \|B\|^2 \|A \sqrt{\omega}\|^2 = \sqrt{2} \|A\|^2_\omega \|B\|^2.
$$

\bigskip

\noindent [The case where $B=\ketbra{\lambda_1}{\lambda_n}$]

\bigskip

Our numerical experiments indicate that the optimal matrix $B$ for achieving the bound $\tilde{c_2}(\omega)$, as specified in \eqref{tcopt}, invariably takes the form $B = \ketbra{\lambda_1}{\lambda_n}$ (after the normalization).
Once we assume this fact, we can prove conjecture \eqref{Conjii} as follows. 
Denoting $a_{ij}:= \bracket{\lambda_i}{A \lambda_j}$, a direct computation yields $\norm{[A,B]}_\omega^2=\lambda_n\sum_{j}|a_{j1}|^2-2\lambda_n\Re (a_{nn}\overline{a_{11}}) + \sum_j\lambda_j|a_{nj}|^2$ , $\norm{A}_\omega^2=\sum_j\lambda_j\sum_k|a_{kj}|^2$, $\|B\|^2=1$. 
Consequently, it follows that 
\begin{align}
	&(\lm+\lM)\norm{A}^2_\omega\norm{B}^2 - \lm\norm{[A,B]}_\omega^2\nonumber\\
	=& \lambda_1\sum_{\substack{j \\ k\ne n}}\lambda_j|a_{kj}|^2 + \lambda_n\sum_{\substack{j\ne 1 \\ k}}\lambda_j|a_{kj}|^2 + 2\lambda_1\lambda_n\Re ( a_{nn}\overline{a_{11}}) \\ 
	=&\Bigl|\lambda_1a_{11}+\lambda_na_{nn}\Bigr|^2 + \lambda_1^2 \sum_{\substack{k\ne1,n}}|a_{k1}|^2 + \lambda_1\sum_{\substack{j\ne1 \\ k\ne n}}\lambda_j|a_{kj}|^2 + \lambda^2_n\sum_{\substack{k \neq n}}|a_{kn}|^2 + \lambda_n\sum_{\substack{j\ne 1,n \\ k}}\lambda_j|a_{kj}|^2 \ge 0
\end{align}

\bibliographystyle{elsarticle-num}
\bibliography{OmegaNormRef}

\begin{thebibliography}{10}
\expandafter\ifx\csname url\endcsname\relax
  \def\url#1{\texttt{#1}}\fi
\expandafter\ifx\csname urlprefix\endcsname\relax\def\urlprefix{URL }\fi
\expandafter\ifx\csname href\endcsname\relax
  \def\href#1#2{#2} \def\path#1{#1}\fi

\bibitem{BOTTCHER2005216}
A.~Böttcher, D.~Wenzel, How big can the commutator of two matrices be and how
  big is it typically?, Linear Algebra and its Applications 403 (2005)
  216--228.

\bibitem{vong2008proof}
S.-W. Vong, X.-Q. Jin, Proof of {B{\"o}ttcher} and {Wenzel}’s conjecture,
  Oper. Matrices 2~(3) (2008) 435--442.

\bibitem{BOTTCHER20081864}
A.~Böttcher, D.~Wenzel, The {Frobenius} norm and the commutator, Linear
  Algebra and its Applications 429~(8) (2008) 1864--1885.

\bibitem{AUDENAERT20101126}
K.~M. Audenaert, Variance bounds, with an application to norm bounds for
  commutators, Linear Algebra and its Applications 432~(5) (2010) 1126--1143.

\bibitem{WENZEL20101726}
D.~Wenzel, K.~M. Audenaert, Impressions of convexity: An illustration for
  commutator bounds, Linear Algebra and its Applications 433~(11) (2010)
  1726--1759.

\bibitem{CHENG2015409}
C.-M. Cheng, C.~Lei, On {Schatten} $p$-norms of commutators, Linear Algebra and
  its Applications 484 (2015) 409--434.

\bibitem{CHRUSCINSKI202295}
D.~Chruściński, G.~Kimura, H.~Ohno, T.~Singal, Bounding the {Frobenius} norm
  of a $q$-deformed commutator, Linear Algebra and its Applications 646 (2022)
  95--106.

\bibitem{CHRUSCINSKI2023158}
D.~Chruściński, G.~Kimura, H.~Ohno, T.~Singal, One-parameter generalization
  of the {B\"ottcher-Wenzel} inequality and its application to open quantum
  dynamics, Linear Algebra and its Applications 656 (2023) 158--166.

\bibitem{Hayashi2015}
M.~Hayashi, S.~Ishizaka, A.~Kawachi, G.~Kimura, T.~Ogawa, Introduction to
  Quantum Information Science, 2015th Edition, Graduate Texts in Physics,
  Springer, Berlin, Heidelberg, 2015.

\bibitem{Mayumi_2023}
A.~Mayumi, G.~Kimura, H.~Ohno, D.~Chruściński, In preparation (2024).

\bibitem{Gorini1976CompletelyPositive}
V.~Gorini, A.~Kossakowski, E.~Sudarshan, Completely positive dynamical
  semigroups of {$N$}‐level systems, Journal of Mathematical Physics 17~(5)
  (1976) 821--825.

\bibitem{Lindblad1976OnTheGenerators}
G.~Lindblad, On the generators of quantum dynamical semigroups, Communications
  in Mathematical Physics 48~(2) (1976) 119--130.

\bibitem{Chruscinski2017BriefHistoryGKLS}
D.~Chruściński, S.~Pascazio, A brief history of the {GKLS} equation, Open
  Systems \& Information Dynamics 24~(03) (2017) 1740001.

\bibitem{PhysRevLett.127.050401}
D.~Chruściński, G.~Kimura, A.~Kossakowski, Y.~Shishido, Universal constraint
  for relaxation rates for quantum dynamical semigroup, Phys. Rev. Lett. 127
  (2021) 050401.

\bibitem{WolfCirac2008}
M.~M. Wolf, I.~Cirac, Dividing quantum channels, Communications in Mathematical
  Physics 279 (2008) 147.

\bibitem{KimuraAjisakaWatanabe2017}
G.~Kimura, S.~Ajisaka, K.~Watanabe, Universal constraints on relaxation times
  for $d$-level {GKLS} master equations, Open Systems \& Information Dynamics
  24~(4) (2017) 1--8.

\end{thebibliography}

\end{document}